# When Good Harmony Is Not Enough: Preserving Human Intention in Explainable Music Generation

**Josef Pavlíček[1], Petra Pavlíčková[2] and Irena Štrausová[1]**

[1] Faculty of Economics, Department of Applied Mathematics and Informatics, University of South Bohemia in České Budějovice (Czech Republic)

[2] Faculty of Information Technology, Czech Technical University in Prague (Czech Republic)

**ABSTRACT**

Although automatic harmonization can produce musically plausible chord progressions, technical correctness does not necessarily ensure that the resulting music reflects human expectations. This study investigates whether explicit guidance toward harmonic stability improves perceived quality within the HARMONIX framework. Six melodic excerpts were rendered with three harmonizations: the original human-annotated reference (HUMAN), the corrected Explainable Melody Harmonizer (BASELINE), and a guided version applying a fixed chord-change penalty of 0.15 (GUIDED). In a randomized, browser-based listening experiment, 57 participants evaluated naturalness, melodic fit, and agreement with their preference for harmonic activity, yielding 1,026 complete variant evaluations. The BASELINE and HUMAN conditions received similar mean ratings, whereas GUIDED was rated consistently lower across all three dimensions. Paired comparisons at participant level, with Holm correction, confirmed the differences between GUIDED and BASELINE. The findings suggest that reducing harmonic changes through a uniform penalty does not necessarily improve the listening experience. They highlight the need for context-sensitive guidance and for distinguishing measurable structural properties from musical intention, which cannot be observed directly.

## INTRODUCTION

Automatic music generation has become an established field of artificial intelligence, spanning rule-based composition, statistical modeling, and data-driven approaches capable of producing melodies, chord progressions, and complete arrangements (Herremans et al., 2018; Briot et al., 2017). Despite these advances, the ability to generate a formally plausible musical sequence does not guarantee that listeners will experience it as natural or expressive.

Musical quality is difficult to reduce to a single computational objective because the meaning of a harmonic choice depends on tonal context, learned expectations, and the relationship between local events and larger musical structures (Lerdahl and Jackendoff, 1983; Krumhansl, 2001; Chew, 2014). A chord can be compatible with the current melody while changing the perceived direction or character of a passage, making the distinction between technical fit and human musical judgment particularly important.

Automatic harmonization offers a focused setting in which to investigate this distinction. Given the same melody, several chord sequences may be defensible, yet they can differ substantially in harmonic rhythm, continuity, and expressive effect. Selecting among these alternatives is therefore not merely a matter of finding valid chords; it also involves decisions about when harmonic change is appropriate (Paiement et al., 2006; Moss et al., 2019).

Our previous research has explored explainable harmonic visualization through BPMN, the Chord Wheel Diagram, and eye-tracking studies (Pavlicek et al., 2024; Pavlicek et al., 2025; Pavlíček et al., 2026c), alongside computational approaches to generating and comparing harmonic alternatives (Pavlíček et al., 2026a; Pavlíček, 2026). These studies raise a further question: can transparent harmonic decisions also preserve the structural qualities that listeners consider meaningful?

To examine this issue empirically, we use the HARMONIX framework to compare original human-annotated harmonies with two computational alternatives. The BASELINE condition employs the corrected Explainable Melody Harmonizer, whereas the GUIDED condition adds a fixed preference for harmonic persistence. Rather than assuming that fewer chord changes are inherently beneficial, we evaluate how this intervention affects listener judgments.

The central objective is to determine whether explicit structural guidance aligns with perceived naturalness, melodic compatibility, and stated preferences for harmonic activity. A randomized listening study allows these outcomes to be compared within the same excerpts and participants, while keeping the underlying melody and rendering instrument constant.

## BACKGROUND AND RESEARCH QUESTION

### Automatic harmonization

Automatic harmonization assigns a sequence of chords to a melody by evaluating candidate harmonies, their compatibility with melodic material, and transitions between successive segments. Both rule-based and learned approaches must balance local fit against progression-level coherence (Paiement et al., 2006; Herremans et al., 2018). Because a melody rarely specifies a unique harmonic solution, the selected sequence can influence musical expectations even when every individual chord is theoretically acceptable (Chew, 2014).

### Explainable harmonic decisions

Explainable AI aims to make computational decisions understandable by exposing their criteria, relationships, and alternatives (Gunning et al., 2019; Adadi and Berrada, 2018; Barredo Arrieta et al., 2020). In music, tonal-harmony visualization provides an additional means of interpreting such decisions (Bunks et al., 2022), while explainable harmonization can connect chord choices with harmonic function, tonal proximity, and competing alternatives.

Our earlier work used BPMN and the Chord Wheel Diagram to represent harmonic development (Pavlicek et al., 2024; Pavlicek et al., 2025). Later work examined the use of these visual methods in music education (Pavlíček et al., 2026b). HARMONIX extends this research with computational harmonization and explicit structural constraints.

### Research questions

We investigate three complementary research questions:

**RQ1:** How does explicit guidance toward harmonic stability change the structure of generated chord sequences?

**RQ2:** How do listeners evaluate the naturalness and melodic fit of human, baseline, and guided harmonizations?

**RQ3:** Does a listener's stated preference for harmonic stability relate to their evaluation of the generated harmony?

## STRUCTURALLY GUIDED HARMONIZATION

### Research framework

HARMONIX generates chord progressions for monophonic melodies by dividing the input into harmonic segments and evaluating candidate chords in relation to the melodic material and adjacent harmonic decisions. The corrected Explainable Melody Harmonizer provides an explicit scoring model, while sequence-level optimization selects a progression that balances local compatibility with continuity across segments.

### Baseline and guided generation

BASELINE uses the corrected Explainable Melody Harmonizer scoring model without the experimental chord-change penalty. It provides a computational comparison that preserves the standard candidate evaluation and sequence-selection procedure.

GUIDED retains the same generation procedure but introduces a fixed penalty of 0.15 for chord changes. This intervention makes repeated chords relatively more favorable when competing solutions have otherwise similar scores, thereby encouraging harmonic persistence without forbidding transitions.

The parameter was fixed before the listening experiment and was not fitted to participant ratings. Accordingly, the study tests one reproducible configuration rather than a personalized or adaptively optimized form of guidance.

### Human reference

HUMAN consists of the annotated harmonic progression associated with each source excerpt and is evaluated independently of the generated alternatives.

The HUMAN annotations were not used as targets to generate the BASELINE or GUIDED outputs. They serve as a documented comparison rather than as the only musically correct solution.

### Scope of human guidance

Although the wider HARMONIX framework can represent repetition, harmonic function, and tonal relationships, the present experiment isolates one intervention: a uniform preference for harmonic persistence. We therefore do not claim to reconstruct an original composer’s intention; instead, we test the perceptual consequences of an explicitly defined structural constraint.

## STUDY MATERIAL AND METHOD

### Musical material

Six short melodic excerpts were selected from HookTheory-derived research material. Each excerpt was paired with the original annotated harmony (HUMAN), the corrected EMH output (BASELINE), and the persistence-guided output (GUIDED). All versions preserve the same melody and use a consistent piano rendering to limit timbral differences between conditions. The six excerpts lasted between 24.2 and 31.4 seconds (M = 27.0 s), with identical durations across the three harmonization conditions for each excerpt.

- HUMAN: the annotated reference harmony.
- BASELINE: harmony generated by the corrected EMH model.
- GUIDED: harmony generated with an additional chord-change penalty.

The audio examples were prepared before data collection and presented without identifying their generation condition, reducing the possibility that labels such as “human” or “AI” would influence judgments.

### Listening experiment

Participants completed a browser-based listening task designed primarily for mobile devices. Both the order of the six excerpts and the assignment of the three harmonization conditions to neutral labels A, B, and C were randomized, while each participant evaluated all three versions of every excerpt.

Before listening, participants indicated whether they generally preferred calmer harmony with fewer changes, more varied harmony with frequent changes, or had no clear preference. This single question captured a stated preference for harmonic activity rather than a comprehensive measure of musical taste.

- **I prefer calm harmony with fewer chord changes.**
- **I prefer more varied harmony with frequent chord changes.**
- **I have no clear preference.**

### Evaluation questions

The evaluation covered three complementary perceptual dimensions:

Q1 (Naturalness): How natural and pleasant does this music sound to you?
Q2 (Melody fit): How well do the accompanying chords fit the melody?
Q3 (Preference match): How well does the number of chord changes match your preference?

All responses used a five-point ordinal scale, with higher values indicating a more favorable evaluation. The questions were phrased without specialist terminology so that formal musical training was not required.

### Participants and data collection

The completed dataset contains 57 anonymous respondents, each of whom evaluated all 18 audio variants, producing 1,026 complete variant evaluations and 3,078 individual scale ratings. Participation was voluntary, and the recorded fields included the harmonization condition, excerpt, ratings, stated preference, and playback-related measures.

The study deliberately focused on everyday music listeners rather than professional musicians, reflecting the broad audience for AI-generated music. Of the 57 participants, 34 (59.6%) reported no musical training, 13 (22.8%) basic training, and 10 (17.5%) informal musical experience. Age was optionally reported by 25 participants (18–67 years, median 23). Although the convenience sample is not representative of the general population, its repeated-measures design enables direct within-participant comparisons of harmonization conditions. This approach supports the human-centered objective of evaluating perceived harmonic quality from the perspective of ordinary listeners rather than expert theoretical judgment.

## DATA ANALYSIS

### Structural analysis

Chord changes were counted between consecutive harmonic segments, allowing direct structural comparison of BASELINE and GUIDED under identical segmentation.

### Listener evaluation

For naturalness, melody fit, and preference match, we report condition-level means across all participant–excerpt evaluations. Paired t-tests compare the participant-level condition means, with 95% confidence intervals and Holm correction across nine planned pairwise comparisons. Wilcoxon signed-rank tests provide a nonparametric robustness check.

As an exploratory engagement check, the same condition comparisons were repeated for participants whose average maximum playback position reached at least 80% of audio duration. This threshold was applied as a sensitivity analysis and not as a criterion for removing responses from the primary dataset.

### Individual preferences

We also examine whether participants who initially preferred stable harmony rated the GUIDED condition more favorably than those who preferred varied or neutral harmony. Because subgroup sizes differ, these comparisons are treated as exploratory rather than confirmatory.

The central interpretive question is whether an explicit structural preference produces a perceptually desirable result, or whether the perceived quality of a progression depends more strongly on where harmonic changes occur and how they relate to melodic context.

## RESULTS AND LIMITATIONS

Structural analysis confirmed that the persistence penalty substantially altered the generated chord sequences. Across the six excerpts, BASELINE contained 64 chord changes, compared with only 19 in GUIDED, representing a 70.3% reduction. The reduction occurred in every excerpt, ranging from 44.4% to 100%. In one case, GUIDED retained the same chord across all 16 harmonic segments, whereas BASELINE included five changes. Thus, the intervention produced a pronounced structural effect rather than a minor adjustment to harmonic continuity.

The principal result is that BASELINE received mean ratings close to HUMAN, whereas GUIDED was evaluated less favorably across all three dimensions (Table 1; Figure 1). Participant-level comparisons confirmed significant disadvantages for GUIDED relative to BASELINE (Table 2). These findings concern the tested implementation rather than harmonic stability or human guidance in general.

**Table 1. Mean listener ratings by harmonization condition (1–5; 57 participants, six excerpts).**

| Condition | Naturalness | Melody fit | Preference match |
|---|---|---|---|
| HUMAN | 3.52 | 3.58 | 3.21 |
| BASELINE | 3.52 | 3.51 | 3.24 |
| GUIDED | 2.86 | 3.04 | 2.55 |

**Table 2. Selected participant-level paired differences with 95% confidence intervals and Holm-adjusted p-values (correction across nine comparisons).**

| Outcome | Contrast | Mean Δ | 95% CI | p (Holm) |
|---|---|---|---|---|
| Naturalness | BASELINE − HUMAN | 0.00 | [-0.11, +0.11] | 1.00 |
| Naturalness | GUIDED − BASELINE | -0.65 | [-0.83, -0.48] | < 0.001 |
| Melody Fit | BASELINE − HUMAN | -0.07 | [-0.19, +0.04] | 0.63 |
| Melody Fit | GUIDED − BASELINE | -0.47 | [-0.64, -0.29] | < 0.001 |
| Preference Match | BASELINE − HUMAN | +0.03 | [-0.09, +0.15] | 1.00 |
| Preference Match | GUIDED − BASELINE | -0.69 | [-0.86, -0.52] | < 0.001 |

Across all 57 participants, HUMAN and BASELINE obtained closely similar mean ratings (Table 1), whereas GUIDED was consistently lower. At participant level, the mean difference between GUIDED and BASELINE was −0.65 for naturalness (95% CI [−0.83, −0.48]), −0.47 for melodic fit (95% CI [−0.64, −0.29]), and −0.69 for preference match (95% CI [−0.86, −0.52]); all three contrasts remained significant after Holm correction (Table 2). In contrast,

BASELINE–HUMAN differences were not significant. The same qualitative pattern was retained in an exploratory analysis restricted to 33 participants who reached at least 80% of the audio duration on average. These results indicate a substantial perceptual disadvantage for the tested GUIDED implementation, without establishing statistical equivalence between HUMAN and BASELINE.

An exploratory analysis of stated harmonic preferences revealed a further distinction between structural guidance and perceived musical quality. Among the 57 participants, seven preferred stable harmony, 22 preferred more varied harmony, and 28 expressed no clear preference. Participants favoring stability rated GUIDED higher on preference match ($M = 3.12$) than those favoring variation ($M = 2.36$) or expressing no preference ($M = 2.55$). Nevertheless, even within the stability-preferring group, GUIDED received lower ratings than HUMAN ($M = 3.55$) and BASELINE ($M = 3.52$). These descriptive findings suggest that a general preference for fewer chord changes does not necessarily translate into a preference for harmonizations generated with a uniform persistence penalty.

Several limitations qualify the interpretation:

- First, recruitment was nonprobabilistic, and the participant pool cannot be assumed to represent all music listeners. Musical experience and listening conditions may also vary across respondents.
- Second, the six short excerpts cover only a limited range of musical contexts, and listeners may evaluate isolated progressions differently from complete compositions.
- Third, GUIDED used one fixed chord-change penalty (0.15) rather than an individualized or context-dependent preference model; consequently, the findings cannot identify an optimal guidance strategy.
- Fourth, the HUMAN condition is one annotated reference for each excerpt, not a unique ground truth for musical correctness or intention.

Finally, the study observes listener ratings and measurable playback behavior but does not directly observe the source composer's intentions. Further structural analyses across a wider range of musical excerpts, together with larger and more diverse listening studies, are needed to establish the generality of these findings.

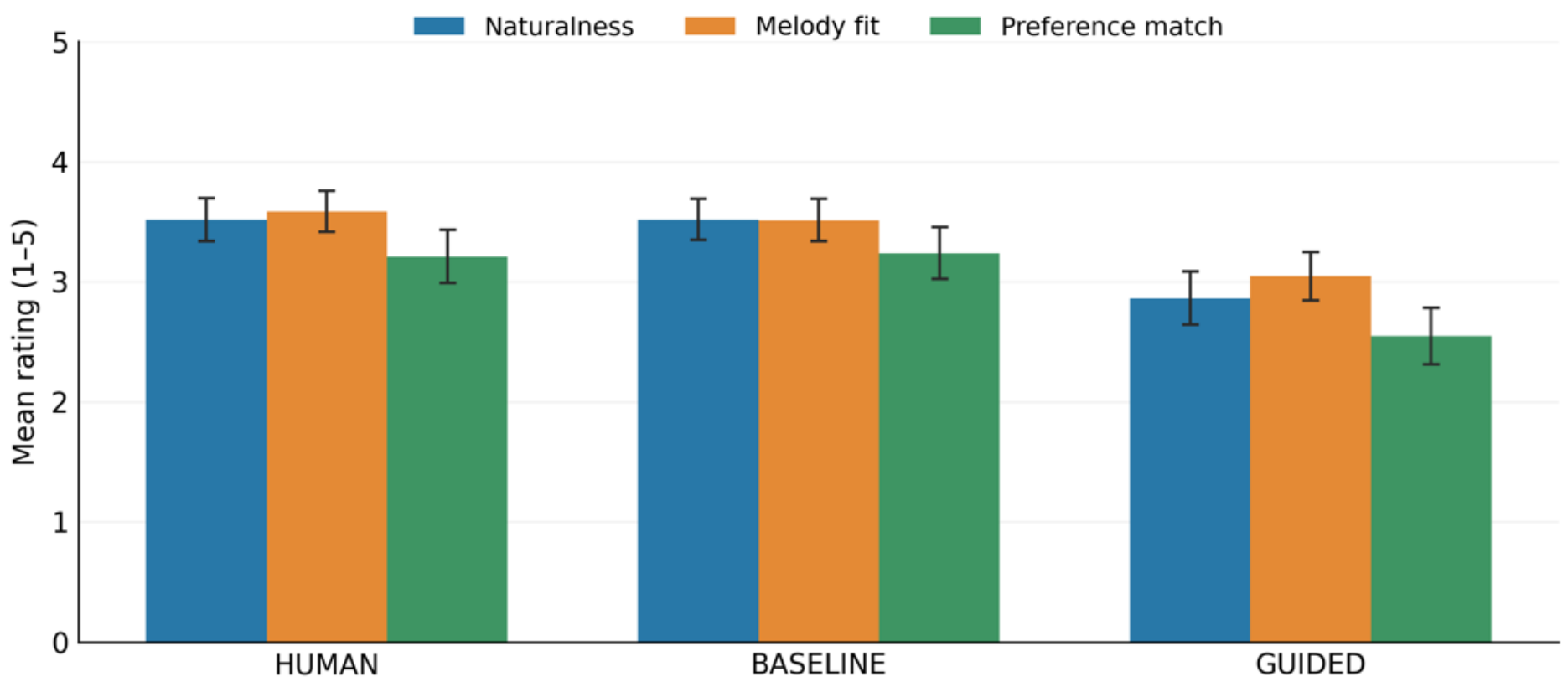

**Figure 1. Mean listener ratings across the three harmonization conditions (1**

**5 scale; error bars represent 95% confidence intervals based on 57 participants).**

## DISCUSSION

The listening results reveal a meaningful distinction between the ability to generate a plausible harmonic progression and the ability to improve it through a simple structural constraint. BASELINE and HUMAN received closely similar average evaluations, suggesting that the corrected computational model can produce harmonizations that listeners judge favorably within the selected excerpts. The absence of a significant difference, however, should not be interpreted as formal statistical equivalence.

By contrast, GUIDED received consistently lower ratings, even though its design explicitly favored harmonic persistence. This result is compatible with the view that musical preference depends on the interaction between stability, expectation, and harmonic surprise rather than on the number of changes alone (Miles et al., 2017). Repetition may support coherence, but an appropriate transition may also provide direction, tension, or resolution.

The observed disadvantage of GUIDED therefore points to a limitation of uniform guidance rather than to a limitation of human-centered generation itself. A fixed penalty does not distinguish between a musically redundant change and a change required by the melodic or functional context, so its effect may conflict with the listener's broader perception of phrase structure.

This interpretation connects the experiment with work on constrained generative systems, where explicit constraints can make output properties controllable but cannot guarantee that the resulting artifact will be preferred (Bonlarron et al., 2026). The current study adds a listening-based evaluation of that trade-off rather than assessing structural compliance alone.

The three rating dimensions help clarify this point. Naturalness, melodic fit, and agreement with a preference for harmonic activity were assessed separately, and all three favored BASELINE over GUIDED. A participant may state a preference for fewer changes yet still dislike a specific progression with repeated chords; therefore, explicit preference elicitation should not be confused with a complete model of musical intention.

Explainable AI can make these decisions inspectable (Barredo Arrieta et al., 2020), but meaningful human agency also requires mechanisms that translate preferences into context-sensitive outcomes. The findings motivate a distinction between explaining why a chord was selected and ensuring that a user-specified constraint is applied in a musically appropriate way.

In its current form, HARMONIX implements a controlled, reproducible intervention rather than an interactive personalization system. The experiment establishes a baseline for investigating more expressive guidance mechanisms, including constraints tied to phrase boundaries, cadences, and harmonic function.

Future work should compare multiple penalty strengths and allow listeners to adjust harmonic stability interactively, ideally through constraints sensitive to phrase boundaries and harmonic function. Such a design would help determine whether preference-sensitive guidance can outperform a fixed global penalty.

Longer musical contexts and larger participant samples would further clarify whether the effects observed here generalize beyond the six selected excerpts and whether user preferences change after hearing alternative harmonizations.

## CONCLUSION

A computationally plausible harmony is not necessarily the harmony a listener prefers, and an explicit preference for stability does not automatically produce a more satisfying musical result.

This paper evaluated three harmonization conditions within the explainable HARMONIX framework using six melodic excerpts and a randomized listening study with 57 respondents. The GUIDED condition reduced chord changes by 70.3% compared with BASELINE but received significantly lower ratings for naturalness, melodic fit, and preference match. Meanwhile, the corrected computational BASELINE received ratings comparable to the HUMAN reference, although statistical equivalence was not established.

The findings demonstrate the importance of distinguishing structural control from subjective musical quality. Although a persistence constraint can be defined and implemented transparently, its perceptual benefit depends on musical context and cannot be assumed from the constraint alone. A substantial reduction in harmonic changes does not necessarily correspond to an improvement in the listening experience.

These results support further development of context-sensitive and interactive guidance in explainable music generation, particularly approaches that consider melodic phrasing, harmonic function, and individual preferences. Future studies should examine different levels of structural guidance and broader musical contexts to determine how human preferences can meaningfully influence computational harmonic decisions.

## ACKNOWLEDGMENT

The authors acknowledge the contributors to the HookTheory dataset and its distribution through the SheetSage project. The HARMONIX listening-study implementation, experimental procedures, analysis scripts, and reproducibility materials are available in the IHSI2027-listening-study directory of the GitHub repository:

https://github.com/JosefPavlicek/quantum-inspired-music-research

Dataset access and reuse remain subject to the original licensing and distribution terms. Generative AI tools assisted with English-language editing and graphical preparation. The authors retain full responsibility for the scientific content, methodology, results, and conclusions.